# Reacting Polymers with Highly Correlated Initial Conditions

OLEG V. BYCHUK[1,2], BEN O'SHAUGHNESSY[1a]
and NICHOLAS J. TURRO[2]

[1]Department of Chemistry,
Columbia University, New York, NY 10027

[2]Department of Chemical Engineering,
Columbia University, New York, NY 10027, USA

PACS: 82.35.+t, 82.40.-g

# ABSTRACT

We propose and theoretically study an experiment designed to measure short time polymer reaction kinetics in melts or dilute solutions. The photolysis of groups centrally located along chain backbones, one group per chain, creates pairs of spatially highly correlated macroradicals. We calculate time-dependent rate coefficients $\kappa(t)$ governing their first order recombination kinetics, which are novel on account of the far-from-equilibrium initial conditions. In dilute solutions (good solvents) reaction kinetics are intrinsically weak, despite the highly reactive radical groups involved. This leads to a generalised mean field kinetics in which the rate of radical density decay $-\dot{n} \sim S(t)$ where $S(t) \sim t^{-(1+g/3)}$ is the *equilibrium* return probability for 2 reactive groups, given initial contact. Here $g \approx 0.27$ is the correlation hole exponent for self-avoiding chain ends. For times beyond the longest coil relaxation time $\tau$, $-\dot{n} \sim S(t)$ remains true, but center of gravity coil diffusion takes over with rms displacement of reactive groups $x(t) \sim t^{1/2}$ and $S(t) \sim 1/x^3(t)$. At the shortest times ($t \lesssim 10^{-6}$ sec), recombination is inhibited due to spin selection rules and we find $\dot{n} \sim tS(t)$. In melts, kinetics are intrinsically diffusion-controlled, leading to entirely different rate laws. During the regime limited by spin selection rules, the density of radicals decays linearly, $n(0) - n(t) \sim t$. At longer times the standard result $-\dot{n} \sim dx^3(t)/dt$ (for randomly distributed ends) is replaced by $\dot{n} \sim d^2x^3(t)/dt^2$ for these correlated initial conditions. The long time behavior, $t > \tau$, has the same scaling form in time as for dilute solutions.

## I. Introduction

For small molecules in the liquid phase reaction kinetics follow the "law of mass action," i.e. the reaction rate is proportional to the *equilibrium contact probability*. For two reactive species A and B this translates into the product of their mean number densities, $n_A$ and $n_B$, with the coefficient of proportionality $k$ being the second order rate constant:

$$\dot{n}_A = \dot{n}_B = -k\, n_A n_B \ . \tag{1}$$

More generally, we term kinetics such that rates are proportional to the equilibrium contact probability of the reactive species "mean-field" (MF) kinetics [1]. For instance, for reacting polymer systems of the type shown in fig. 1, where the reactive species A and B are attached to polymer chain ends, mean field kinetics would imply a rate proportional to the equilibrium probability for the ends of two polymer chains to occupy the same position in space. The rate constant $k$ would then depend not only on chemical characteristics of the functional groups, but also on properties of the host polymer coils such as chain length $N$ and polymer volume fraction $\phi$ [1]. For example, in dilute and semi-dilute polymer solutions contact of the chain ends (which necessitates the A and B coils to strongly interpenetrate one another) can only be achieved by overcoming interchain excluded volume repulsions. As will be discussed below, the MF class of kinetics is not the only one. A second general class is "diffusion-controlled" (DC) kinetics. In the case of polymers, reaction kinetics are sometimes of MF type and sometimes of DC type, depending on the concentration regime and local reactivity of the reactive groups.

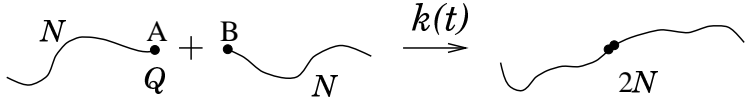

Figure 1: Reactive species A and B attached to polymer chain ends. Two reactive chains, each comprising $N$ monomer units, form a chemically inert chain of length $2N$. If A and B are within reaction range $b$ of each other, reaction occurs with probability $Q$ per unit time. For highly reactive species, e.g. radicals, $Q$ is limited by diffusion on scale $b$, such that $Q \approx t_b^{-1}$ where $t_b$ is the monomer relaxation time.

In fact, theoretical studies predict that polymer reaction kinetics are rather novel: for times short compared to the longest polymer relaxation time $\tau$ the rate "constant" $k$ may actually be *time-dependent* [2–4]. This interesting time dependence in $k(t)$ and, more generally, the short time behavior of $k$ for reacting polymers, is the subject of this paper. In cases where $k(t)$ does depend on time, this derives from the fact that for short times kinetics are DC: the reaction rate is limited by how rapidly diffusion can transport reactive groups into contact. In contrast, when hydrodynamical and excluded volume interactions are strong, it has been predicted theoretically [1] that, even for highly reactive groups, MF kinetics are recovered with a time-independent rate coefficient $k$ even at small times. Whether MF or DC kinetics apply, in all cases $k(t)$ reflects fundamental properties of the host polymer chains, and its measurement probes fundamental properties of polymer melts and solutions.

Here, we will study solutions and melts of polymeric chains with one reactive group (a radical) per chain end; see fig. 2. Chains react irreversibly by radical recombination.

The resulting chains, twice the length of reactants, are chemically inert. Reactions of this type arise, for example, in free radical polymerization, where they serve as the principle termination mechanism for growing "living" chains [5–15].

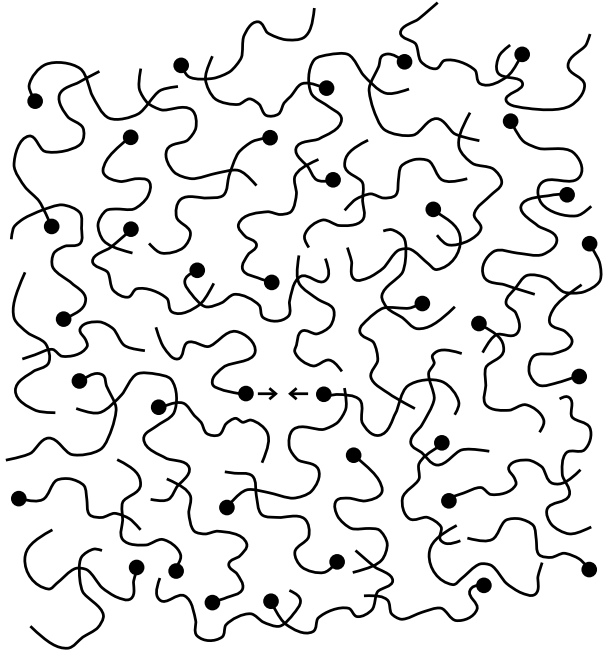

Figure 2: End-functionalized polymer chains ($N$ monomer units, coil radius $R$), either dispersed within a polymer melt of inert chains or dissolved in a good small-molecular-weight solvent. The initial density of reactive groups is $n_0$.

Unfortunately, no experimental measurements of $k(t)$ in polymeric systems have been reported to date, to our knowledge. Given its fundamental importance this may seem odd. However, there is a profound experimental obstacle: the fraction of neighboring reactive chain ends in equilibrium is very small. For example, in a melt of chains of $N$ units each, the probability for an end to be in contact with another end is approximately equal to the very small volume fraction of chain ends $N^{-1}$. This renders the short-time signal in most conventional methods (e.g. phosphorescence quenching [16,17]) too weak to measure.

In this paper we present the theoretical analysis of an experiment which we propose can measure the short-time behavior of reaction rates by avoiding this low-signal problem. The idea (see fig. 3) is to create reactive free radical ends in pairs by cleaving polymer chains at the central location along their backbones by a flash of laser light; fig. 4 shows a candidate molecule. Following the photolysis, each radical has exactly one nearby radical with which to recombine. The recombination takes place with probability $Q$ per unit time if the radicals are within reactive range $b$ of each other. For highly reactive chemical species, such as radicals, reactions essentially occur instantaneously [18] on contact (i.e. they react on a timescale shorter than or of the same order as the monomer diffusion time $t_b \approx 10^{-10}\,\text{sec}$). Thus effectively $Q \approx t_b^{-1}$ [19]. This situation will always pertain in the present paper.

In order to simplify the interpretation, the incoming light intensity and/or the concentration of photocleavable chains will be chosen sufficiently low such that recombinations between different macroradical pairs can be neglected due to the low concentration of radicals that are produced. The contact probability, and therefore the reaction rate, per radical is thus independent of the radical number density $n$; the reaction rate per unit volume, $\dot{n}$, is therefore proportional to $n$. These are first order kinetics characterized by a first order rate "constant" $\kappa(t)$:

$$\dot{n} = -\kappa(t)\, n \ . \qquad (2)$$

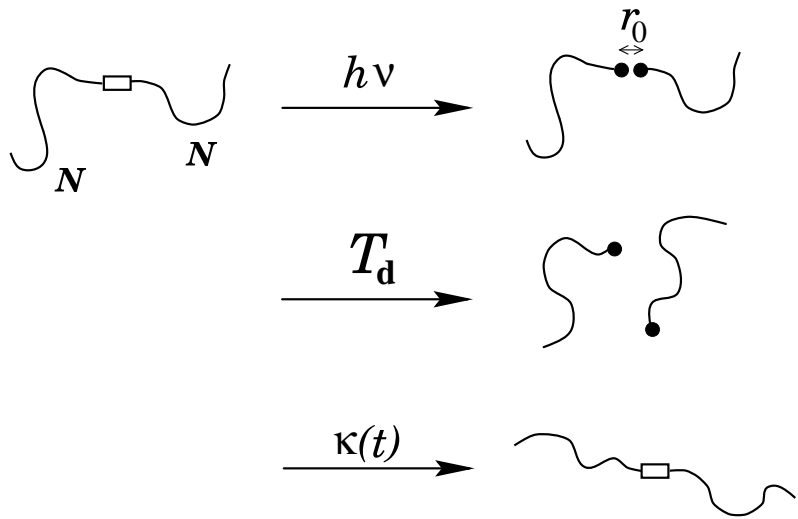

Figure 3: Schematic of the proposed photocleaving experiment. A laser pulse cleaves a labeled chain, producing two macroradicals separated by distance $r_0$. Immediate recombination is prohibited by spin selection rules. After a typical delay time $T_d$ recombination becomes possible. Reaction kinetics are thereafter characterized by a time-dependent first-order rate "constant" $\kappa(t)$.

In fact, as we will demonstrate later, a more precise statement of the above is that $\dot{n} = -\kappa(t)\, n$ is true in MF cases only. In situations where DC kinetics apply, $\dot{n}$ is proportional to the *initial* number density $n_0$ and the definition of $\kappa$ is slightly different: $\dot{n} = -\kappa(t)\, n(0)$. This "memory" of initial conditions results from the fact that reactions in DC cases present strong perturbations to polymer kinetics.

Note that $\kappa(t)$ has dimensions of inverse time, in contrast to $k(t)$, whose "volume over time" dimensions reflect second order kinetics. The number of unreacted radicals at any moment can be measured [18,20–22] by techniques such as light absorption in the ultraviolet, visible or infrared range or electron paramagnetic resonance (EPR). The main experimental advantage of this technique is that the experimental signal is proportional to the number density of radicals $n$. By contrast, in conventional methods [16,17] the signal is proportional to the equilibrium number of pairs in contact, i. e. to $n^2$. This is a much weaker signal.

Of course, immediate recombination of radicals following the cleavage might preclude any measurement of $\kappa(t)$. It turns out that this is prevented from happening by spin selection rules [18]. The two macroradicals are formed in a spin 1 (triplet) pair [18] whose recombination is prohibited. After a characteristic time delay $T_d$ a transition ("intersystem crossing"; see ref. 18) to a spin 0 (singlet) state occurs, and recombination becomes allowed. During $T_{\rm d}$ the radical ends will diffuse apart, and when the prohibition is removed the reactive ends are well separated; their subsequent approach is governed by polymer dynamics.

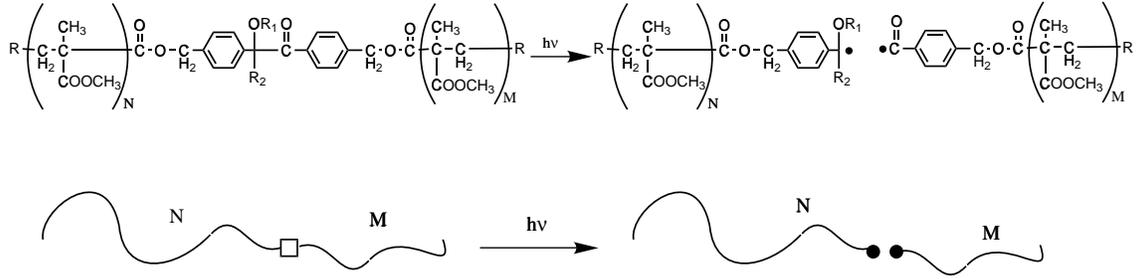

Figure 4: An example of a photocleavable polymer chain, polymethylmethacrylate with an incorporated photolyzable aromatic ketoether [18]. We consider M=N in this paper.

We develop the theory of this original technique and predict a range of unusual and interesting time-dependent behaviors, reflecting DC or MF kinetics relevant for different systems. In section II we review standard polymer reaction kinetics, where chains are distributed as in equilibrium at the moment $t = 0$ when the reactions are "switched on." In sections III and IV we present the theory of reaction kinetics for the case of the highly non-equilibrium initial distribution pertaining to the proposed experiment. The results are summarized and discussed in section V.

## II. Two Classes of Reaction Kinetics: Mean-Field and Diffusion-Controlled

A time-dependent reaction rate "constant" may seem peculiar. But in fact it arises even in the simple problem of reacting Brownian particles in one spatial dimension [23–26]. If the local reactivity $Q$, i. e. the rate of reaction per unit time when the reactive groups overlap, is sufficiently high, then for short times the number of reactions per given particle after time $t$ is equal to the total number of other particles within diffusive range [4], namely $n_0 x_t$, where $n_0$ is the initial number density and $x_t$ is the root mean square (rms) displacement of a molecule after time $t$. The reaction rate per unit length will then be $\dot{n}(t) \equiv -k(t) n_0^2 \approx -n_0 \, {\rm d}(n_0 x_t)/{\rm d}t$ where $n(t)$ is the number density at time $t$; hence $k(t) \approx {\rm d}x_t/{\rm d}t$. For Brownian diffusion $x_t \sim t^{1/2}$, and hence the rate constant decays with time: $k(t) \sim t^{-1/2}$.

In polymer systems $x_t$, the rms displacement of a reactive group attached to a polymer chain, can scale with various powers of $t$ for short times $t < \tau$. In order to see how this leads to various forms of $k(t)$, consider all pairs of reactive groups which are initially within $x_t$ of each other. At $t = 0$ reactions commence. After time $t$ the separation $\mathbf{r}$ of two such groups will be confined to a region of volume of order $x_t^3$. The distribution of $\mathbf{r}$ within $x_t^3$ will be approximately [27] the equilibrium distribution *given* the condition that $\mathbf{r}$ is confined to $x_t^3$. In melts excluded volume effects are screened [28], $\mathbf{r}$ is distributed approximately uniformly within $x_t^3$ and thus the contact probability (the probability $r \leq b$) scales like $1/x_t^3$. Hence the number of collisions after time $t$ scales as $t/x_t^3$, which in melts is an increasing function of time since $x_t \sim t^\beta$ where $\beta \leq 1/4$ in all cases [29]. That is, for any finite $Q$ value, reaction for such a pair is certain for sufficiently large times. For such times almost all pairs initially within diffusive range $x_t$ of each other will have reacted: these are DC kinetics. The number of such pairs grows as $x_t^3$ with time. Therefore [2–4] $k(t) \approx dx_t^3/dt$. As an example, for unentangled melts $x_t \sim t^{1/4}$ [28] so that the number of collisions grows as $t^{1/4}$ and $k(t) \sim t^{-1/4}$ [4].

In good solvents excluded volume repulsions reduce the contact probability by a factor proportional to $1/x_t^g$ where the "correlation hole" exponent $g \approx 0.27$ for chain ends [30,31]. Therefore the mean number of collisions decays with time as [19] $t/x_t^{(3+g)} \sim t^{-g/3}$ (in good solvents $x_t \sim t^{1/3}$ [28]). That is, only a small fraction of the reactive group pairs starting within diffusive range $x_t$ will have reacted, and the distribution of reactive groups is only weakly perturbed from equilibrium. Hence MF kinetics apply [27], with rate constant [27,19]

$$k = k_\infty^{\text{dil}} \approx Qb^3 \left(\frac{b}{R}\right)^g \sim \frac{1}{N^{\nu g}} \approx \frac{1}{N^{0.16}} \qquad \text{(dilute; good solvent)} \qquad (3)$$

where $\nu \approx 3/5$ is the Flory exponent [28] defining the rms coil size, $R \approx bN^\nu$ ($b$ is the monomer size). The rate constant for all times equals the long time MF value, $k_\infty^{\text{dil}}$.

Long time reaction kinetics in melts are different. For $t > \tau$ monomer displacement is determined by the Fickian diffusion of the polymer coil as a whole, i. e. $x_t \sim t^{1/2}$, and kinetics must cross over from DC to MF. Demanding continuity at $t \approx \tau$, the long time $k$ must equal its small time value evaluated at $t \approx \tau$ [2–4]. Since $x_\tau \approx R$, this gives $k \approx R^3/\tau$.

The above theoretical predictions are essentially untested experimentally, aside from evidence of weak molecular weight dependence of the long time $k$ in good solvents from phosphorescence quenching experiments [16]. Our proposed experiment, described in the introduction, is expected to overcome the weak signals of past efforts. However, there is a profound difference between this experiment and the systems studied theoretically in the past, in that the reactive groups are not initially randomly distributed over space, but instead start from a highly correlated non-equilibrium initial state.

## III. Reaction Kinetics: Highly Correlated Initial Distribution

In the previous section we discussed DC and MF kinetic regimes in polymer reactions. There we considered the situation where the initial location of reactive groups was as in equilibrium. For end-functionalized chains this means very low contact probability for reactive groups.

In the photolysis experiment proposed in this paper the initial state is highly non-equilibrium: reactive macroradicals are arranged in pairs, each pair forming an intertwined coil (recall that before the laser flash each pair was a single chain). More importantly, the reactive end groups are in close proximity. This leads to a fundamental difference between the proposed experiment and the usual situation discussed in section II: The reaction rate is now proportional to the number of macroradical pairs, and therefore to the number of reactive groups $n(t)$. These first order kinetics are characterized by a first order rate constant $\kappa(t)$ defined in the introduction.

How does $\kappa(t)$ relate to the standard second-order rate constant $k(t)$? Can the photolysis experiment probe the theoretically well-studied reaction kinetics for the case of equilibrium initial distribution? In this and the following sections we will calculate the evolution of $n(t)$ in the photolysis experiment and see how and when it can be related to the standard second order kinetics.

For simplicity, in this section we will consider the case when the "forbidden" period $T_d$ during which recombinations are prohibited is very short, specifically of order the monomer diffusion time $t_b$. When the reactions are switched on, the initial separation of the reactive groups $r_0$ is thus of order the reaction radius $b$. The case of general $T_d$ will be treated in section IV.

Let us consider some general properties of the dynamics of a reactive pair. It is convenient to introduce the *survival probability* $\Pi(t)$, namely the fraction of the initially photolyzed polymers which have not recombined by time $t$:

$$\Pi(t) \equiv \frac{n(t)}{n_0} \ . \tag{4}$$

For a given radical pair $\Pi(t)$ is the probability that the pair has remained unreacted by time $t$.

The dynamics of $\Pi$ are closely related to the dynamics of $p(\mathbf{r}, t)$, the probability density for the reactive groups to be separated by $\mathbf{r}$ at time $t$. In fact, by its definition $\Pi$ is the (decaying) normalization of $p$:

$$\Pi(t) = \int d\mathbf{r} \ p(\mathbf{r}, t) \ . \tag{5}$$

At the same time $\Pi$ must depend only on the probability for the reactive groups to be in contact,

$$p_{\text{cont}}(t) \equiv \int_{r<b} d\mathbf{r} \ p(\mathbf{r}, t) \tag{6}$$

according to the relation

$$1 - \Pi(t) = Q \int_0^t dt' \ p_{\text{cont}}(t') \ . \tag{7}$$

This relationship states that the probability $1 - \Pi(t)$ for a pair to have recombined by time $t$ is the local reactivity $Q$ times the mean time the reactive groups have spent in contact. Note that $p_{\text{cont}}(t) Q dt$ is the probability to react between times $t$ and $t + dt$. Eq. (7) follows from from the dynamical equation for $p$, see Appendix A.

In this section we consider the special case where $r_0$ is of the same order as the reaction radius, $r_0 \approx b$. In appendix A we show, starting from the dynamics of $p$, that for such an initial condition the Laplace transform $(t \to E)$ of the survival probability is given by

$$\Pi_E \approx \frac{1}{E} \frac{1 + Q \left[ S_E - \overline{S}_{\widetilde{E}} \right]}{1 + Q S_E} \qquad (r_0 = b; \ Et_b < 1) \ . \tag{8}$$

Here $S_E$ is the Laplace transform of the non-reactive return probability [4,27], namely the probability for two monomers to overlap (i.e. to lie within $b$ of each other) at time $t$ given they were initially in contact. $\overline{S}_{\widetilde{E}}$ is the same return probability as $S_E$, but now it is given that the initial monomer separation was $r_0$, with $r_0$ different to but of the same order as $b$.

In consequence, for small $E$ the difference $S_E - \overline{S}_{\widetilde{E}}$ is contributed to from small times of order $t_b$ only. In appendix B this difference is shown to be of order $t_b$ for $Et_b \ll 1$.

The return probability can be expressed in terms of the Green's function $G_t(\mathbf{r}, \mathbf{r}_0)$, i. e. the probability density, in the absence of reactions, for two monomers to be separated by $\mathbf{r}$ after time $t$ given initial separation $\mathbf{r}_0$ [4,27]:

$$S(t) \equiv \int_{r<b} d\mathbf{r} \, G_t(\mathbf{r}, 0) \; . \tag{9}$$

Note the quite general property that $S(0) = 1$. In fact $S(t) \approx 1$ for any $t \lesssim t_b$. All power law decays in $S$ derived below are understood to be cut off at $t_b$ in this fashion.

**(i) Melts.** $S(t)$ in melts can be found by the following argument. If two monomers start together, their separation after time $t$ is less than or of order $x_t$. Since excluded volume interactions in melts are screened [28], the distribution of their separations $\mathbf{r}$ within the volume $x_t^3$ is approximately uniform. Hence [4,27]

$$S(t) \approx \left(\frac{b}{x_t}\right)^3 \qquad (t > t_b; \text{ melt}) \; . \tag{10}$$

The rms displacement $x_t$ scales with time with various exponents, $x_t \sim t^\beta$. In the unentangled case [29] for $t < \tau = N^2 t_b$ the exponent is $\beta = 1/4$. For longer times Fickian diffusion is recovered with $\beta = 1/2$. When entanglements are present, separated on average by $N_e$ monomer units along the backbone, the reptation theory [29] predicts four distinct regimes separated by the following time scales: $\tau_e \equiv t_b N_e^2$, the relaxation time of a chain segment between two neighboring entanglements; $\tau_R = \tau_e (N/N_e)^2$, the Rouse time; and the longest coil relaxation time $\tau = \tau_{\text{rep}} \approx \tau_e (N/N_e)^3$, the time scale on which a chain reptates clear of all its original entanglements. The values of $\beta$ corresponding to these regimes are $1/4$, $1/8$, $1/4$ and $1/2$ respectively [29]. Note that the integral of the short-time ($t < \tau$) value of $S(t)$ diverges at long times ($\beta \leq 1/4$ in all cases): the number of collisions between two monomers grows without bound as time progresses. As discussed in section II, this is related to the DC nature of kinetics in melts.

In appendix B we calculate $S_E$ and $\overline{S}_{\widetilde{E}}$ and show $S_E - \overline{S} \approx t_b$ and $QS_E \gg 1$ for $Et_b \ll 1$. Hence the expression (8) for the survival probability becomes

$$\Pi_E \approx \frac{1}{\widetilde{Q}ES_E} \; , \qquad \widetilde{Q} = \frac{Q}{1 + Qt_b} \; . \tag{11}$$

Note that the effective reactivity $\widetilde{Q}$ saturates for large $Q$ at the monomer diffusion rate, $\widetilde{Q} \to t_b^{-1}$ for $Q \gg 1/t_b$. Now we can evaluate $\Pi(t)$ with the help in the case of highly reactive radicals ($Qt_b \gg 1$) the effective reactivity is given by the diffusion time on scale $t_b$: $\widetilde{Q} \approx t_b^{-1}$. Now we can evaluate $\Pi(t)$ with the help of the following property of Laplace transforms [32] ($\mathcal{L}$ signifies Laplace transform): $E\mathcal{L}[t^\alpha] \approx (E\mathcal{L}[t^{-\alpha}])^{-1}$ for $0 < \alpha < 1$. Since $x_t$ is a power law, from eqs. (10) and (11) for $Et_b < 1$ one has $\Pi_E \approx \left(\widetilde{Q}b^3 E\mathcal{L}[x_t^{-3}]\right)^{-1} \approx \left(\widetilde{Q}b^3\right)^{-1} E\mathcal{L}[x_t^3]$, and

$$\Pi(t) \approx \left(\widetilde{Q}b^3\right)^{-1} \frac{d}{dt} x_t^3 \qquad (t < \tau, \text{ melt}) \; . \tag{12}$$

This means that the reaction rate is proportional to the initial radical number density $n_0$ (see definition of $\Pi$, eq. (4)), and eq. (12) defines the DC reaction rate coefficient $\kappa^{\text{DC}}(t)$:

$$\dot{n}(t) = -\kappa^{\text{DC}}(t)\, n_0\,, \quad \kappa^{\text{DC}}(t) \approx \left(\widetilde{Q} b^3\right)^{-1} \frac{\mathrm{d}^2}{\mathrm{d}t^2} x_t^3 \qquad (t < \tau,\ \text{melt})\,. \tag{13}$$

For unentangled melts $x_t \sim t^{1/4}$ for $t < \tau$, so

$$\Pi(t) \approx \frac{1}{\widetilde{Q} t_b} \left(\frac{t_b}{t}\right)^{1/4}\,, \qquad \kappa^{\text{DC}}(t) \sim t^{-5/4} \qquad (t < \tau;\ \text{unentangled melt})\,. \tag{14}$$

Note that a fraction of order unity of radicals have recombined by time $t_b$. This reflects the singular nature of the $r_0 = b$ situation in melts: DC kinetics apply and most reactions occur on the short diffusive time scale $t_b$; see section II.

In the entangled case

$$\widetilde{Q} t_b\, \Pi(t) \approx \begin{cases} (t_b/t)^{1/4} & (t < \tau_e) \\ N_e^{-1/2}\, (\tau_e/t)^{5/8} & (\tau_e < t < \tau_R) \\ N_e^{-1/2}\, (N_e/N)^{5/4}\, (\tau_R/t)^{1/4} & (\tau_R < t < \tau_{\text{rep}}) \end{cases} \qquad \text{(entangled melt)} \tag{15}$$

and

$$\kappa^{\text{DC}}(t) \sim t^{-\zeta}\,, \qquad \zeta = \begin{cases} 5/4 & (t < \tau_e) \\ 13/8 & (\tau_e < t < \tau_R) \\ 5/4 & (\tau_R < t < \tau_{\text{rep}}) \end{cases} \qquad \text{(entangled melt)}\,. \tag{16}$$

For times longer than $\tau$ two unreacted radical-bearing coils drift apart obeying Fickian diffusion, $x_t \approx R(t/\tau)^{1/2}$, but they remain spatially correlated: they are much closer to each other than to other pairs. The integral of the return probability now in fact converges (see appendix B), reflecting the MF nature of kinetics for such long times:

$$S(t) \approx S(\tau) \left(\frac{\tau}{t}\right)^{3/2}\,, \qquad S_E \approx \tau S(\tau)\left[1 - (E\tau)^{1/2}\right] \qquad (t > \tau;\ \text{melt})\,. \tag{17}$$

Substituting this into the expression for $\Pi_E$, eq. (11), yields, after inverse Laplace-transforming,

$$\Pi(t) \approx \Pi_\infty \left[1 + \left(\frac{\tau}{t}\right)^{1/2}\right]\,, \quad \Pi_\infty \approx \frac{1}{\widetilde{Q} \tau S(\tau)}\,, \quad \kappa^{\text{MF}}(t) \approx \frac{1}{\tau}\left(\frac{\tau}{t}\right)^{3/2} \quad (t > \tau;\ \text{melt})\,. \tag{18}$$

Here we used the fact that in melts $\widetilde{Q}\tau S(\tau) \gg 1$. Indeed, in the unentangled case $\widetilde{Q}\tau S(\tau) \approx \widetilde{Q}\tau(b/R)^3 \approx \widetilde{Q} t_b N^{1/2}$ (recall $\widetilde{Q} \approx t_b^{-1}$ for radicals); in the presence of entanglements [29] $\widetilde{Q}\tau S(\tau) \approx \widetilde{Q} t_b N^{3/2}/N_e$. For times longer than $t_l$, the diffusion time to the nearest neighbor radical pair which lies a distance $\tilde{n}_0^{-1/3}$ away, the non-equilibrium initial distribution is forgotten and standard MF reaction kinetics are recovered. Here $\tilde{n}_0 \approx \Pi_\infty n_0$ is an effective initial density for the onset of 2nd order kinetics. Indeed, setting $t = t_l$ in the above expression for $\kappa^{\mathrm{MF}}(t)$, one obtains $\kappa^{\mathrm{MF}}(t_l) \approx R^3 \tilde{n}_0/\tau$, crossing over to to the standard 2nd order kinetics with rate constant $k \approx R^3/\tau$; see introduction.

**(ii) Dilute solutions, good solvent.** In good solvent conditions excluded volume repulsions play a crucial role. After time $t$, on length scales less than $x_t$ the chains have had time to equilibrate. The distribution of $\mathbf{r}$ values ($\mathbf{r}$ being the radical pair separation) is thus essentially the conditional equilibrium distribution *given* that $r < x_t$. Thus the probability of separation less than $r$ is approximately $(r/x_t)^{3+g}$ [1,19]. The return probability equals this expression evaluated at $r = b$:

$$S(t) \approx \left(\frac{b}{x_t}\right)^{3+g} \approx \begin{cases} \left(\frac{t_b}{t}\right)^{1+g/3} & (t_b < t < \tau) \\ \left(\frac{b}{R}\right)^{3+g} \left(\frac{\tau}{t}\right)^{3/2} & (t > \tau) \end{cases}. \tag{19}$$

This result was derived in refs. 1. Excluded volume repulsions reduce the return probability relative to the ideal melts result (eq. (10)) by the factor $(b/x_t)^g$ [1]. This corresponds to the equilibrium correlation hole [30,31] mentioned in section II

Notice that, unlike the melts case for $t < \tau$, the time integral of $S(t)$ now converges: that is, in the absence of reactions the chain ends do not collide with certainty. The total probability of collision is approximately $Q \int_0^\infty dt\, S(t) \approx Q t_b$ (recall $Q \approx t_b^{-1}$); correspondingly, a finite fraction of radical pairs escape reaction altogether. The rapid decay of $S(t)$ means that for $t > t_b$ reactions only weakly perturb what the distribution of chain end separations would be were there no reactions.

The Laplace transforms $S_E$ and $\overline{S}_{\widetilde{E}}$ are calculated in appendix B. Substitution into eq. (8) gives

$$\Pi_E \approx \begin{cases} \dfrac{1}{E} \dfrac{1}{\alpha - \widetilde{Q} t_b (E t_b)^{g/3}} & (E\tau > 1) \\[2mm] \dfrac{1}{E} \dfrac{1}{\alpha - \widetilde{Q}\tau S(\tau)\left[1 - (E\tau)^{1/2}\right]} & (E\tau < 1) \end{cases} \tag{20}$$

where $\alpha \approx (1 + AQt_b)/(1 + BQt_b)$, with $A$ and $B$ constants of order unity, is itself a constant of order unity ($\alpha > 1$). $\widetilde{Q}$ is defined in eq. (11). After inverse Laplace-transforming, this becomes

$$\frac{\Pi(t)}{\Pi_\infty} \approx \begin{cases} 1 + \widetilde{Q} t_b \left(\dfrac{t_b}{t}\right)^{g/3} & (t < \tau) \\[2mm] 1 + \widetilde{Q}\tau S(\tau) \left(\dfrac{\tau}{t}\right)^{1/2} & (t > \tau) \end{cases} \tag{21}$$

with $\Pi_\infty \approx 1/\alpha$. We see that a fraction of order unity of radicals escape unreacted.

For $t > \tau$, as in the case of melts, two unreacted coils diffuse apart with $x_t \approx R(t/\tau)^{1/2}$ while remaining spatially correlated. From eq. (21)

$$\kappa^{\mathrm{MF}}(t) \approx Q \left(\frac{b}{R}\right)^{3+g} \left(\frac{\tau}{t}\right)^{3/2} = \frac{k_\infty^{\mathrm{dil}}}{(Dt)^{3/2}} \qquad (\tau < t < t_l; \text{ good solvent}) \tag{22}$$

where $k_\infty^{\rm dil}$ is the dilute solution 2nd order rate constant for standard reaction kinetics (eq. (3)) and $D \approx R^2/\tau \sim 1/R$ is the long time translational diffusion coefficient [28]. For times longer than the diffusion time to the nearest neighbor radical pair $t_l$, the standard MF reaction kinetics are recovered. Indeed, setting $t = t_l$ in the above expression, one sees $\kappa^{\rm MF}(t_l) \approx k_\infty^{\rm dil} \widetilde{n}_0$ with $\widetilde{n}_0 \approx \Pi_\infty n_0$, which correctly crosses over to the standard 2nd order kinetics with rate constant $k_\infty^{\rm dil}$.

## IV. The Effect of Delayed Recombination

In the previous section we assumed that recombination was allowed almost immediately after photocleaving, more precisely after a short delay of order the monomer time scale $t_b$. In reality there is a longer delay before recombinations start due to spin selection rules, as discussed in the introduction. We assume here that for any given radical pair there exists a unique time $T_{\rm d}$ after which they can recombine if they come into contact. For $t < T_{\rm d}$ recombination is strictly prohibited. (This simplified picture ignores the coupling between spin state and spatial dynamics, as mentioned in introduction.)

In the actual experiment there is of course an ensemble of radical pairs and a corresponding distribution of delay times, $P(T_{\rm d})$, with a certain characteristic width $T_{\rm d}^0$ such that $P(T_{\rm d})$ is exponentially small for $T_{\rm d} > T_{\rm d}^0$. We begin, however, with the conceptually important case of a single $T_{\rm d}$ value, i.e. $P$ is imagined to be sharply peaked at one value.

### A. Sharply Peaked Distribution of $T_{\rm d}$

During the time interval $T_{\rm d}$ following photocleaving the newly-created radicals diffuse without reacting. Even though the initial conditions are highly non-equilibrium, by the time reactions begin the chains will have had time to equilibrate on length scales smaller than $L_{\rm d} \equiv x_{T_{\rm d}}$. Therefore the radical separation $\mathbf{r}_0$ at $t = T_{\rm d}$ is governed for $r_0 < L_{\rm d}$ approximately by the conditional equilibrium pair distribution, the condition being that the radicals are within $L_{\rm d}$ of each other [27,19].

In this subsection it will be convenient to phrase calculations in terms of a shifted time coordinate, $\widetilde{t} \equiv t - T_{\rm d}$. In appendix C we calculate $\Pi_{\widetilde{E}}$, the Laplace transform ($\widetilde{t} \to \widetilde{E}$) of $\Pi(\widetilde{t})$, the survival probability a time $\widetilde{t}$ after the recombination prohibition has been removed. We show that $\Pi_{\widetilde{E}}$ is related to $S_{\widetilde{E}}$, the Laplace transform ($\widetilde{t} \to \widetilde{E}$) of $S(\widetilde{t})$, and $\overline{S}_{\widetilde{E}}$, the Laplace transform of

$$\overline{S}\left(\widetilde{t}\right) \equiv S(t) , \tag{23}$$

by an expression formally identical to eq. (8). (Note that the function $\overline{S}$ employed for the case of no delay, section III, can now be identified as the special case of eq. (23) when $T_{\rm d} = t_b$.)

Noting that $\overline{S}$ is just the original return probability but shifted in time, $\overline{S}(\widetilde{t}) = S(\widetilde{t} + T_{\rm d})$, thus $\overline{S}_{\widetilde{E}}$ can be written in the following useful form:

$$\overline{S}_{\widetilde{E}} = e^{\widetilde{E}T_{\rm d}} \int_{T_{\rm d}}^\infty d\widetilde{t}\, e^{-\widetilde{E}\widetilde{t}} S\left(\widetilde{t}\right) . \tag{24}$$

**(i) Melts.** Integration by parts of eq. (24) gives for $\tilde{E}T_{\rm d} > 1$

$$\overline{S}_{\tilde{E}} = \frac{1}{\tilde{E}} S(T_{\rm d}) \left[ 1 + O\left(\frac{1}{\tilde{E}T_{\rm d}}\right) \right] \qquad (\tilde{E}T_{\rm d} > 1) \ . \qquad (25)$$

For $\tilde{E}T_{\rm d} < 1$ the first exponential on the right hand side of eq. (24) gives corrections of relative order $\tilde{E}T_{\rm d}$. Therefore it can be omitted with the result

$$\overline{S}_{\tilde{E}} \approx S_{\tilde{E}} - T_{\rm d} \frac{b^3}{\Omega} \qquad (\tilde{E}T_{\rm d} < 1) \qquad (26)$$

where we used the expression for $S(t)$, eq. (10) with $x_t \sim t^\beta$, and $\Omega \equiv L_{\rm d}^3$. Substitution of eqs. (25) and (26) into eq. (8) gives

$$\Pi_{\tilde{E}} \approx \begin{cases} \dfrac{1}{\tilde{E}} \left[ 1 - \dfrac{ST_{\rm d}}{(\tilde{E}t_b)^{3\beta}} \right] & (\tilde{E}T_{\rm d} > 1) \\[2ex] T_{\rm d} \dfrac{b^3}{\Omega} \dfrac{1}{\tilde{E}\overline{S}_{\tilde{E}}} & (\tilde{E}T_{\rm d} < 1) \end{cases} \qquad (27)$$

where we used $Q\overline{S}_{\tilde{E}} \gg 1$ for $\tilde{E}t_b \ll 1$ (see eq. (B4); recall that $Q \approx 1/t_b$ for radicals). Substituting $S_{\tilde{E}}$ from appendix B, eq. (B4), one obtains, after inverse Laplace-transforming:

$$\Pi\left(\tilde{t}\right) \approx \begin{cases} 1 - \dfrac{x_{\tilde{t}}^3}{\Omega} & (\tilde{t} < T_{\rm d}) \\[2ex] \left(\dfrac{T_{\rm d}}{\tilde{t}}\right)^{1-3\beta} & (T_{\rm d} < \tilde{t} < \tau) \\[2ex] \left(\dfrac{T_{\rm d}}{\tau}\right)^{1-3\beta} \left[ 1 + \left(\dfrac{\tau}{\tilde{t}}\right)^{1/2} \right] & (\tilde{t} > \tau) \end{cases} \qquad (28)$$

**(ii) Dilute solutions, good solvent.** For $\tilde{E}T_{\rm d} > 1$ the expression for $\overline{S}_{\tilde{E}}$ in eq. (25) is still valid. For $\tilde{E}T_{\rm d} < 1$, integration of eq. (24) gives

$$\overline{S}_{\tilde{E}} \approx \begin{cases} T_{\rm d} S(T_{\rm d}) \left[ 1 - (\tilde{E}T_{\rm d})^{9/3} \right] & (\tilde{E}\tau > 1) \\[1ex] T_{\rm d} S(T_{\rm d}) - \tau S(\tau)(\tilde{E}\tau)^{1/2} & (\tilde{E}\tau < 1) \end{cases} \ . \qquad (29)$$

We observe that in all three cases the time dependence in the denominator of the expression for $\Pi_{\tilde{E}}$, eq. (8), leads to a higher order correction. Therefore we can write

$$\Pi_{\tilde{E}} \approx \frac{1}{\tilde{E}} \left[ 1 - \tilde{Q}\,\overline{S}_{\tilde{E}} \right] \qquad (30)$$

with $\tilde{Q}$ given in eq. (11). Substituting $S_{\tilde{E}}$ and inverse Laplace-transforming, one obtains

$$\Pi(\tilde{t}) \approx \begin{cases} 1 - \tilde{Q}S(T_d)\tilde{t} & (\tilde{t} < T_d) \\ \Pi_\infty + \tilde{Q}T_d S(T_d) \left(\dfrac{T_d}{\tilde{t}}\right)^{g/3} & (T_d < \tilde{t} < \tau) \\ \Pi_\infty + \tilde{Q}\tau S(\tau) \left(\dfrac{\tau}{\tilde{t}}\right)^{g/3} & (\tilde{t} > \tau) \end{cases}, \qquad \tilde{t} \equiv t - T_d \qquad (31)$$

with the asymptotic unreacted fraction given by

$$\Pi_\infty \approx 1 - \tilde{Q}T_d S(T_d) \ . \qquad (32)$$

## B. Broadly Distributed $T_d$

In the actual experiment the fraction of radical pairs which undergo a transition to the singlet state between $T_d$ and $T_d + dT_d$ after photocleaving equals $P(T_d)dT_d$. Of these a fraction $R(t|T_d) \equiv 1 - \Pi(t|T_d)$ will have recombined by a later time $t$; here $\Pi(t|T_d)$ is the conditional survival probability after time $t$ given the recombination becomes allowed at time $T_d$. Hence the observed fraction of reacted radicals is

$$R(t) = \int_0^t dT_d \ P(T_d) \ R(t|T_d) \ . \qquad (33)$$

**(i) Melts.** In this case for $t < \tau$ from eq. (28) $R(t|T_d)$ reduces to a function of one argument, $w \equiv (t - T_d)/T_d$:

$$R(t|T_d) = f(w) \ , \qquad f(w) \approx \begin{cases} w^{3\beta} & (w < 1) \\ 1 - w^{-(1-3\beta)} & (w > 1) \end{cases} . \qquad (34)$$

Then, after a change of variables,

$$R(t) = t \int_0^\infty \frac{dw}{(1+w)^2} \ P\left(\frac{t}{1+w}\right) \ f(w) \approx \begin{cases} \dfrac{t}{T_d^0} & (t < T_d^0) \\ 1 - \left(\dfrac{T_d^0}{t}\right)^{1-3\beta} & (T_d^0 < t < \tau) \end{cases} \qquad \text{(melt)} \qquad (35)$$

where we used the fact that by the normalization condition $P(T_d) \approx (T_d^0)^{-1}$ for $T_d \lesssim T_d^0$. For $t > \tau$ one obtains, after using $\int_0^t dT_d P(T_d) \approx 1$,

$$\Pi(t) \approx \left(\frac{T_d^0}{\tau}\right)^{1-3\beta} \left[1 + \left(\frac{\tau}{t}\right)^{1/2}\right] \qquad (t > \tau; \text{ melt}) \ . \qquad (36)$$

**(ii) Dilute solutions, good solvent.**

For $t < T_d^0$ one can replace $P(T_d)$ with $(T_d^0)^{-1}$ (see eq. (31)). Then, after performing the integration,

$$\Pi(t) \approx 1 - \widetilde{Q} t_b \frac{t_b}{T_d} \left(\frac{t}{t_b}\right)^{1-g/3} \quad (t < T_d^0;\ \text{dilute}) \ . \tag{37}$$

For longer times, only the asymptotic value $\Pi_\infty$ depends on $T_d$. The averaging is straightforward after one replaces the upper limit of integration with $T_d^0$ and $P(T_d)$ for $T_d \lesssim T_d^0$ with $(T_d^0)^{-1}$. Then

$$\Pi(t) \approx \begin{cases} \Pi_\infty + \widetilde{Q} t_b \, (t_b/t)^{g/3} & (T_d^0 < t < \tau) \\ \Pi_\infty + \widetilde{Q} \tau S(\tau) \, (\tau/t)^{1/2} & (t > \tau) \end{cases},$$

$$\Pi_\infty \approx 1 - \widetilde{Q} t_b \left(\frac{t_b}{T_d^0}\right)^{g/3} \ . \tag{38}$$

The general rule for $t > T_d^0$ is that $\Pi(t) \approx \Pi(t|T_d^0)$.

## V. Summary

We have proposed and theoretically studied an experiment designed to measure time-dependent reaction kinetics in liquid polymer systems. The behaviors for melts and dilute solutions are very different.

In the proposed experiment radical pairs are created by photocleaving centrally located groups on the polymer chains. We calculated the radical survival probability, $\Pi(t)$, proportional to the density of unreacted radicals $n(t)$: $\Pi(t) = n(t)/n(0)$. An essential feature is the delay time before a given radical pair can recombine, due to spin selection rules. The delay times are randomly distributed with width $T_d^0 \approx 10^{-8}$ sec [18] in typical cases. In general, we found that after averaging over delay times, the net long time reaction kinetics ($t > T_d^0$) are as is if there were only a single delay time, namely $T_d^0$.

For melts we found that a fraction of order unity of initially present radicals recombines by time $T_d^0$; virtually all of the remaining unreacted radicals recombine between $T_d^0$ and the longest polymer relaxation time $\tau$. This is true provided $T_d^0 \ll \tau$ as is usually the case. Of the remaining few unreacted radicals, for times greater than $\tau$ a fraction of order unity react. Assuming $T_d^0$ is much greater than the monomer diffusion time $t_b \approx 10^{-10}$ sec our

main predictions are:

$$\Pi(t) \approx \begin{cases} 1 - C\dfrac{t}{T_\mathrm{d}^0} & (t < T_\mathrm{d}^0) \\[1ex] \left(\dfrac{T_\mathrm{d}^0}{t}\right)^{1-3\beta} & (T_\mathrm{d}^0 < t < \tau) \qquad \text{(melt)} \\[1ex] \left(\dfrac{T_\mathrm{d}^0}{\tau}\right)^{1/4}\left[1 + \left(\dfrac{\tau}{t}\right)^{1/2}\right] & (t > \tau) \end{cases} \quad (39)$$

Here $C$ is a constant of order unity. For unentangled melts the diffusion exponent $\beta = 1/4$ and $\tau = N^2 t_b$, giving $\tau \approx 10^{-6}$ sec for a typical number of chain units $N \approx 100$. For unentangled melts $\tau = (N^3/N_\mathrm{e})t_b$ is the much longer reptation time, $\tau \approx 10^{-3}$ sec for typical [28,29] values $N = 1000$, $N_\mathrm{e} = 100$. The exponent $\beta$ has successive values $1/4, 1/8, 1/4$ depending on the sub-regime during the interval $0 < t < \tau$.

An experimentally desirable effect of the delay phenomenon, evident in eq. (39), is an *increase* the measured quantity, $n(t) = n_0 \Pi(t)$: the prefactors multiplying the long time power-law decays are proportional to a positive power $T_\mathrm{d}^0$. The bigger the delay, the bigger the signal.

These results for the surviving fraction $\Pi(t) \equiv n(t)/n(0)$ define a sequence of time-dependent first-order reaction rate coefficients, $\dot{n} = -\kappa^\mathrm{DC}(t)\, n(0)$, as

$$\kappa^\mathrm{DC}(t) \sim t^{-\zeta}, \qquad \zeta = \begin{cases} 0 & (t < T_\mathrm{d}^0) \\ 2 - 3\beta & (T_\mathrm{d}^0 < t < \tau) \\ 3/2 & (t > \tau) \end{cases}. \quad (40)$$

These are the same exponents, in fact, as would be observed were there no reaction delay at all (see eq. (16)). The only difference is that the smallest time regimes, $t < T_\mathrm{d}^0$, are absent.

Turning now to dilute solutions, we found that virtually all reactions occur much earlier than $T_\mathrm{d}^0$. The total reacted fraction is very small, reflecting strong excluded volume repulsions and faster relaxation times in dilute environments where the solvent is good. The principal findings (eqs. (37), (38)) are:-

$$\Pi(t) \approx \begin{cases} 1 - \widetilde{Q} t_b \left(\dfrac{t_b}{T_\mathrm{d}^0}\right)^{0.09} \left(\dfrac{t}{T_\mathrm{d}^0}\right)^{0.91} & (t < T_\mathrm{d}^0) \\[1ex] \Pi_\infty + \widetilde{Q} t_b \left(\dfrac{t_b}{t}\right)^{0.09} & (T_\mathrm{d}^0 < t < \tau) \qquad \text{(dilute)} \\[1ex] \Pi_\infty + \widetilde{Q} t_b \left(\dfrac{t_b}{\tau}\right)^{0.09} \left(\dfrac{\tau}{t}\right)^{1/2} & (t > \tau) \end{cases} \quad (41)$$

with $\widetilde{Q} \equiv Q/(1 + Q t_b)$ and $Q$ the local radical reactivity, usually of order $1/t_b$. The asymptotic reacted fraction is

$$\Pi_\infty \approx 1 - \widetilde{Q} t_b \left(\dfrac{t_b}{T_\mathrm{d}^0}\right)^{0.09}. \quad (42)$$

This defines a sequence of time-dependent rate coefficients $\dot{n} = -\kappa^{\mathrm{MF}}(t)\, n$, as

$$\kappa^{\mathrm{MF}}(t) \sim t^{-\zeta}, \qquad \zeta = \begin{cases} 0.09 & (t < T_{\mathrm{d}}^0) \\ 1.09 & (T_{\mathrm{d}}^0 < t < \tau) \\ 3/2 & (t > \tau) \end{cases}. \qquad (43)$$

Note that for the longest times, $t > T_{\mathrm{d}}^0$, the rate coefficient is independent of the characteristic delay time $T_{\mathrm{d}}^0$ which enters only through the asymptotic unreacted fraction, $\Pi_\infty$.

This work was supported by the Petroleum Research Fund under grant no. 33944-AC7, by the National Science Foundation under grant no. NSF-CHE-97-07495 and by the MRSEC Program of the National Science Foundation under Award no. DMR-98-09687. The authors thank Dimitris Vavylonis for many stimulating discussions.

## Appendix A.  Calculation of Survival Probability $\Pi$ without Delay

In this appendix we derive the expression for the survival probability $\Pi$, defined in eq. (4), for the case when the initial separation between the reactive groups, at the moment when reactions are switched on, is of order the reaction range $b$.

From the definition of $\Pi$ it follows that $\Pi(t) = \int d\mathbf{r}\, p(\mathbf{r}, t)$, where $p(\mathbf{r}, t)$ is the probability distribution of the separations $\mathbf{r}$ of reactive groups within a pair at time $t$. It satisfies the following equation involving the non-reactive Green's function $G_t(\mathbf{r}, \mathbf{r}_0)$ [1]:

$$p(\mathbf{r}, t) = \int d\mathbf{r}_0\, G_t(\mathbf{r}, \mathbf{r}_0) p_0(\mathbf{r}_0) - Q \int d\mathbf{r}_0 \int_0^t dt'\, G_{t-t'}(\mathbf{r}, \mathbf{r}_0) \theta(b - r_0) p(\mathbf{r}_0, t') \quad (A1)$$

where $p_0(\mathbf{r}_0) \equiv p(\mathbf{r}_0, 0)$ is the probability distribution of initial separations, and we assume that reactions occur with probability $Q$ per unit time if the reactive groups are within $b$ of each other. The step function $\theta(h)$ equals 1 if $h \geq 0$ and 0 otherwise. Integration over $\mathbf{r}$ gives

$$\Pi(t) = 1 - Q \int_0^t dt'\, p_{\text{cont}}(t') \quad (A2)$$

where we used the fact that $G_t$ and $p_0$ are normalized to unity, and $p_{\text{cont}}(t)$ is the contact probability defined in eq. (6).

To obtain $p_{\text{cont}}(t)$ we integrate the dynamical equation (A1) over the region $r < b$:

$$p_{\text{cont}}(t) = \int_{r<b} d\mathbf{r} \int d\mathbf{r}_0\, G_t(\mathbf{r}, \mathbf{r}_0) p_0(\mathbf{r}_0) - Q \int_{r<b} d\mathbf{r} \int_{r_0<b} d\mathbf{r}_0 \int_0^t dt'\, G_{t-t'}(\mathbf{r}, \mathbf{r}_0) p(\mathbf{r}_0, t') \ . \quad (A3)$$

Since we are considering the case $T_{\text{d}} \approx t_b$, the distribution of initial separations $p_0$ is localized to the region $r_0 \lesssim b$. Therefore, for the times of interest, $t > t_b$, we can set $\mathbf{r}_0 = 0$ in $G_t(\mathbf{r}, \mathbf{r}_0)$ since for such long times $G_t(\mathbf{r}, \mathbf{r}_0)$ depends only very weakly on $\mathbf{r}_0$ for $r_0$ values of order $b$. Then, after using the normalization condition for $p_0(\mathbf{r}_0)$, one obtains

$$p_{\text{cont}}(t) \approx \overline{S}(t) - Q \int_0^t dt'\, S(t - t') p_{\text{cont}}(t') \ . \quad (A4)$$

Here $S(t)$ is the return probability defined in eq. (9). The tilde in the first term indicates that the approximation used here (setting $\mathbf{r}_0 = 0$ in $G_t$) is somewhat different to the approximation in the second term. As we will show, the resulting difference between $\overline{S}$ and $S$ for times less than or of order $t_b$ has important consequences for the long-time behavior of $\Pi(t)$. Now, after Laplace-transforming ($t \to E$), one can solve for $p_{\text{cont}}$:

$$p_{\text{cont}}(E) \approx \frac{\overline{S}_E}{1 + Q S_E} \qquad (E t_b < 1) \ . \quad (A5)$$

Substitution into the Laplace transform of eq. (A2) leads to the expression for $\Pi_E$, eq. (8).

## Appendix B.  Calculation of $S_E$ and $\overline{S}_E$ (No Delay)

In this appendix we calculate the Laplace transforms of the return probability $S(t)$ and the closely related object $\overline{S}(t)$. There are two distinct regimes, $t < \tau$ and $t > \tau$, corresponding to $E\tau > 1$ and $E\tau < 1$ respectively (we are always interested in $t > t_b$, i.e. $Et_b < 1$).

## 1. Melts

For $E\tau > 1$ in the unentangled case ($\beta = 1/4$) using the expression for $S(t)$ for $t > t_b$, eq. (10), one obtains

$$S_E \approx \int_0^{t_b} dt\, S(t) + \int_{t_b}^{\infty} dt \left(\frac{t_b}{t}\right)^{3/4} e^{-Et}$$

$$= \gamma t_b + t_b (E t_b)^{-1/4} \int_0^{\infty} du\, \frac{e^{-u}}{u^{3/4}} + O\left((E t_b)^{1/2}\right)$$

$$\approx t_b (E t_b)^{-1/4} \qquad (E\tau > 1;\ \text{unentangled})\ . \tag{B1}$$

Here we used substitution $u \equiv Et$ and, in the last equality, $E t_b > 1$. The prefactor $\gamma$ is of order unity, and it reflects the details of the small-scale behavior for $t < t_b$. Note that the prefactor in the second term is independent of the small scale details. Thus $\gamma$ is different in $S_E$ and $\overline{S}_E$, while the prefactor in the second term is identical. Hence

$$S_E - \overline{S}_E \approx t_b + O\left((E t_b)^{1/2}\right)\ . \tag{B2}$$

When entanglements are present one can write

$$S_E \approx \int_0^{t_b} dt\, S(t) + \int_{t_b}^{\infty} dt \left(\frac{t_b}{t}\right)^{3\beta} e^{-Et}\ , \tag{B3}$$

$\beta$ now being time-dependent and assuming a sequence of values $1/4$, $1/8$, $1/4$ in the three asymptotic regimes separated by the entanglement time $\tau_e$ and the Rouse time $\tau_R$ (see section III). $S_E$ and $\overline{S}_E$ are each dominated by the relevant value of the exponent $\beta$ at the time $t \approx E^{-1}$, i.e., by analogy with the unentangled case,

$$S_E \approx \overline{S}_E \approx t_b (E t_b)^{-(1-3\beta)}\ . \tag{B4}$$

As in the case of no entanglements, the prefactor multiplying the leading term, i.e. $t_b (E t_b)^{-(1-3\beta)}$, is the same in $S_E$ and $\overline{S}_E$. In fact, this is true for the long-time integral in eq. (B3); this can be seen by writing this integral as a sum of integrals corresponding to the entanglement regimes which happen for times smaller than $E^{-1}$. Therefore eq. (B2) pertains for the entangled case as well.

Consider now long times, corresponding to $E\tau < 1$. We write

$$S_E \approx \overline{S}_E \approx t_b + \int_{t_b}^{\tau} dt \left(\frac{t_b}{t}\right)^{3\beta} e^{-Et} + \left(\frac{b}{R}\right)^3 \int_{\tau}^{\infty} dt \left(\frac{\tau}{t}\right)^{3/2} e^{-Et} \tag{B5}$$

where we used $x_t \sim t^{1/2}$ for $t > \tau$ and

$$S(t) \approx \left(\frac{b}{R}\right)^3 \left(\frac{\tau}{t}\right)^{3/2} \qquad (E\tau < 1;\ \text{melt})\ . \tag{B6}$$

By an argument similar to the one for the $E\tau > 1$ regime the second term in eq. (B5) is independent to leading order of the small scale $t_b$ and, therefore, enters $S_E$ and $\overline{S}_E$ with exactly the same prefactor. The integral in the last term can be calculated by using substitutions $u \equiv t/\tau$ and $s \equiv Et$:

$$\int_\tau^\infty dt \left(\frac{\tau}{t}\right)^{3/2} e^{-Et} = \tau \int_1^\infty \frac{ds}{s^{3/2}} - \tau(E\tau)^{1/2} \int_0^\infty \frac{du\,[1-e^{-u}]}{u^{3/2}} + O(E\tau) \ . \tag{B7}$$

Since the prefactors are explicitly independent of $t_b$, they are the same in $S_E$ and $\overline{S}_E$, and $S_E - \overline{S}_E$ is given by eq. (B2). One can easily check that the above integral dominates $S_E$. Hence

$$S_E \approx \tau S(\tau) \left[1 - (E\tau)^{1/2}\right] \qquad (E\tau < 1;\ \text{melt}) \ . \tag{B8}$$

## 2. Dilute Solution

Using eq. (19), we obtain for $E\tau > 1$:

$$\begin{aligned}
S_E &\approx \int_0^{t_b} dt\, S(t) + \int_{t_b}^\infty dt \left(\frac{t_b}{t}\right)^{1+g/3} e^{-Et} = \gamma t_b + t_b \int_1^\infty ds \frac{e^{-Et_b s}}{s^{1+g/3}} \\
&= \gamma t_b + t_b \int_1^\infty \frac{ds}{s^{1+g/3}} - (Et_b)^{g/3} \int_0^\infty du \frac{1-e^{-u}}{u^{1+g/3}} + O(Et_b) \\
&\approx t_b \left[1 - (Et_b)^{g/3}\right] \qquad (E\tau > 1;\ \text{dilute})
\end{aligned} \tag{B9}$$

where, as in the case of melts, $\gamma$ reflects the details of the small-scale behavior, and we used substitutions $s \equiv t/t_b$ and $u \equiv Et$. Again, the coefficient $\gamma$ depends on small-scale details and is different in $S_E$ and $\overline{S}_E$. The prefactor multiplying $(Et_b)^{g/3}$, on the other hand, does not involve details of the initial condition and, therefore, the corresponding terms in $S_E$ and $\overline{S}_E$ cancel out:

$$S_E - \overline{S}_E \approx t_b + O(Et_b) \ . \tag{B10}$$

For $E\tau < 1$, analogously to the melt case,

$$S_E \approx t_b + \tau S(\tau) \left[1 - (E\tau)^{1/2}\right] \qquad (E\tau < 1;\ \text{dilute}) \ . \tag{B11}$$

## Appendix C. Survival Probability $\Pi$ with Delayed Recombination

Here we calculate $\Pi\left(\tilde{t}\right)$ for the case when recombination is prohibited following photolysis for a time $T_d$ which is the same for all macroradical pairs. Here, as in subsection IV.A, $\tilde{t} \equiv t - T_d$. The function $\overline{S}$ is the return probability shifted in time by an amount $T_d$, eq. (23).

Since the radical pair evolves without recombinations for a time of $T_\text{d}$, the probability distribution of initial separations $\mathbf{r}_0$ when the prohibition is removed is given by the no-reaction Green's function:

$$p_0(\mathbf{r}_0) = G_{T_\text{d}}(\mathbf{r}_0, 0) \ . \tag{C1}$$

Then by the Chapman-Kolmogorov property of Green's functions (see e. g. ref. 33) the first term in the dynamical equation for $p\left(\mathbf{r}, \tilde{t}\right)$, eq. (A1), becomes

$$\int \mathrm{d}\mathbf{r}_0 \ G_{\tilde{t}}(\mathbf{r}, \mathbf{r}_0) p_0(\mathbf{r}_0) = G_{\tilde{t}+T_\text{d}}(\mathbf{r}, 0) \ . \tag{C2}$$

Then integration of eq. (A1) over the region $r < b$ yields, after setting $\mathbf{r}_0 = 0$ in the Green's function in the last term, an equation for the contact probability $p_\text{cont}(\tilde{t})$ analogous to eq. (A4):

$$p_\text{cont}\left(\tilde{t}\right) \approx \overline{S}\left(\tilde{t}\right) - Q \int_0^{\tilde{t}} \mathrm{d}\tilde{t}' \ S\left(\tilde{t} - \tilde{t}'\right) p_\text{cont}\left(\tilde{t}'\right) \ , \quad \overline{S}\left(\tilde{t}\right) \equiv S(\tilde{t} + T_\text{d}) \ . \tag{C3}$$

This form is identical to eq. (A4). Therefore the Laplace transform ($\tilde{t} \to \widetilde{E}$) $\Pi_{\widetilde{E}}$ has the same form in terms of $S_{\widetilde{E}}$ and $\overline{S}_{\widetilde{E}}$, eq. (8), as in the case of $T_\text{d} \approx t_b$.